\documentclass[12pt]{iopart}
\usepackage{graphicx} 
\usepackage{dcolumn}   
\usepackage{bm}        
\usepackage{amssymb}   
\usepackage{amsmath}
\usepackage{color}
\usepackage{amsfonts, amsmath, amssymb}
\usepackage[dvipsnames]{xcolor}
\usepackage{appendix}
\usepackage{booktabs}
\usepackage{url}
\usepackage{hyperref}
\usepackage{tabularx}
\newcommand{\koff}{k_{\mathrm{off}}}
\newcommand{\om}{\omega}
\newcommand{\nkp}{n_{\mathrm{KP}}}
\newcommand{\kp}{k_{\mathrm{p}}}
\newcommand{\Pinc}{\mathcal{P}}
\begin{document}
\title{Kinetic Proofreading Activation Times Beyond the Weibull Limit}

\author{Thomas Mourier$^{1,\dagger}$, Cameron Hopkinson$^2$, \& Kenny Wong$^3$}
\address{$^1$Department of Mechanical and Industrial Engineering, Northeastern University, Boston, MA 02115 \\}
\address{$^2$Department of Biology, Washington University, St. Louis, MO 63130 \\}
\address{$^3$Department of Physics, The Ohio State University, Columbus, Ohio 43210 \\}
\address{$^\dagger$Author to whom any correspondence should be addressed.}
\ead{mourier.t@northeastern.edu}
\date{\today}

\begin{abstract}
T cells can detect a few foreign peptide-major histocompatibility complexes (pMHCs) among a far larger number of self pMHCs. Kinetic proofreading explains this discrimination: a receptor that has bound its ligand must complete $\nkp$ signaling steps, each at rate $\kp$, before it activates, and dissociation at any step resets its progress. When many receptors are engaged at once, the time to the first activation has been approximated by a Weibull law taken from the short-time limit of this process. The Weibull law is convenient, but it does not contain the dissociation rate $\koff$, even though $\koff$ is what proofreading uses to tell ligands apart. Using the exact first-passage law of the underlying continuous-time Markov chain together with stochastic simulations, we show that the relative error of the Weibull law is set by a single dimensionless number $\Lambda$, the Weibull-predicted activation time measured in units of $\nkp+1$ bound-state lifetimes. For $\Lambda\ll1$ the first receptor to activate does so before its ligand has ever dissociated, and the Weibull law holds. As $\Lambda$ grows, activation relies increasingly on repeated unbinding and rebinding. The true activation time then exceeds the Weibull prediction by up to 2900-fold on our parameter grid, and the effective Weibull shape falls to one, so activation becomes memoryless instead of timer-like. We derive a closed form that keeps $\koff$ by treating the successful pass through the proofreading chain separately from the time lost to failed attempts. It matches exact medians to within 2\% in 19 of 25 parameter sets, with a worst case of about 6\%, and reproduces the loss of timer-like behavior. Because $\Lambda$ is computed from the Weibull prediction itself, it shows in advance how far that formula can be trusted; where it cannot, the new expression requires only a one-dimensional root find.
\end{abstract}
\maketitle

\section{Introduction}
T cells decide, through contacts with antigen-presenting cells (APCs), whether any of the peptides displayed on those cells indicate infection or cellular transformation. The peptides are presented by major histocompatibility complex (MHC) molecules, and each APC carries a large population of these peptide-MHC complexes (pMHCs), almost all of them derived from self proteins. T cells can nonetheless respond to very small numbers of foreign pMHCs, in some experiments approaching a single complex~\cite{irvine2002,sykulev1996,purbhoo2004}. Recognition has to be sensitive enough to detect such scarce agonists and specific enough that the far more numerous self pMHCs do not produce comparable responses.
 
Recognition starts when a T cell receptor (TCR) binds a pMHC. In conventional $\alpha\beta$ T cells, the TCR$\alpha$ and TCR$\beta$ chains together recognize the composite surface formed by the peptide and the MHC molecule. These two chains have short cytoplasmic tails and no catalytic activity, so the signal is carried instead by the associated CD3 proteins. The complete receptor has eight chains: the TCR$\alpha\beta$ heterodimer, the CD3$\gamma\varepsilon$ and CD3$\delta\varepsilon$ heterodimers, and a CD3$\zeta\zeta$ homodimer~\cite{smithgarvin2009,courtney2018,gaud2018}. The $\alpha\beta$ chains determine which pMHC is bound, and the CD3 chains relay the binding event into the cell.
 
The CD3 chains carry immunoreceptor tyrosine-based activation motifs (ITAMs), each containing two tyrosines that, once phosphorylated, serve as docking sites for downstream signaling proteins. CD3$\gamma$, CD3$\delta$ and each of the two CD3$\varepsilon$ chains carry one ITAM, and each $\zeta$ chain carries three, giving ten ITAMs and 20 ITAM tyrosines per receptor. None of these chains is a kinase. The catalytic activity comes mainly from tyrosine kinases recruited to, or already positioned near, the receptor~\cite{smithgarvin2009,courtney2018,gaud2018}.
 
One of the first events after productive engagement is phosphorylation of ITAM tyrosines by the Src-family kinase LCK, with a contribution from FYN. LCK associates with the CD4 or CD8 coreceptor, so when the coreceptor binds the same MHC molecule as the TCR, it brings the kinase close to the cytoplasmic tails of the engaged receptor. An ITAM with both tyrosines phosphorylated binds the tandem SH2 domains of ZAP70, a Syk-family kinase. Binding relieves the autoinhibition of ZAP70, and further phosphorylation by LCK, including at sites that activate its kinase domain, makes it catalytically active~\cite{smithgarvin2009,courtney2018}. The first stage of signaling is therefore a sequence of events that must occur in order: pMHC binding, coreceptor engagement, ITAM phosphorylation, and the recruitment and activation of ZAP70.
 
Active ZAP70 phosphorylates the transmembrane adaptor LAT and the cytoplasmic adaptor SLP-76, which together nucleate a larger complex often called the LAT signalosome. Phosphorylated LAT recruits GADS, GRB2 and phospholipase C$\gamma$1 (PLC$\gamma$1), and SLP-76 serves as an additional scaffold for downstream effectors. PLC$\gamma$1 generates the second messengers inositol trisphosphate and diacylglycerol (DAG), which couple proximal tyrosine phosphorylation to calcium mobilization, protein kinase C activity, Ras-MAPK signaling and, eventually, transcriptional programs driven by NFAT, NF-$\kappa$B and AP-1~\cite{smithgarvin2009,courtney2018,gaud2018}. A transient contact at the cell surface can in this way lead to proliferation, cytokine secretion, differentiation or cytotoxic activity.
 
The multistep structure of the cascade matters because equilibrium receptor occupancy does not fully account for antigen discrimination. Ligands that elicit different functional responses can differ strongly in how long they stay bound to the TCR, and early measurements found that T cell responsiveness tracks the dissociation rate $\koff$ of the TCR/pMHC interaction~\cite{matsui1994}. Later work showed that the lifetime a receptor effectively experiences also depends on rebinding and confinement at the cell-cell interface~\cite{aleksic2010}, and that potency depends on ligand affinity more steeply than receptor occupancy alone would predict~\cite{pettmann2021}. These observations call for a kinetic rather than a purely thermodynamic description of antigen recognition.
 
In the 1970s, Hopfield and Ninio independently described a general way for multistep reactions to amplify small differences in molecular lifetime~\cite{hopfield1974,ninio1975}. In kinetic proofreading, one or more effectively irreversible, energy-consuming or delayed steps separate initial recognition from product formation, and at each of them forward progress competes with dissociation. A substrate that stays bound longer gets more chances to advance, while one that dissociates quickly is usually rejected before it reaches the end. A series of such checkpoints turns a modest difference in lifetime into a much larger difference in the probability of completion, which is how high-fidelity biosynthesis can exceed the specificity that equilibrium free-energy differences alone allow~\cite{hopfield1974,ninio1975}.
 
McKeithan adapted the idea to TCR signaling~\cite{mckeithan1995}. In his model, ligand binding puts the receptor in an initial bound state, from which it has to pass through $\nkp$ further states before it activates. Each forward step occurs at the proofreading rate $\kp$, and from every state short of the last the ligand can dissociate at rate $\koff$, returning the receptor to the unbound state and erasing its progress. Since forward motion competes with dissociation at every step, the probability of completing the chain within a single binding event is the per-step success probability raised to the power $\nkp$. A long-lived ligand has a reasonable chance of finishing, and a short-lived one must restart many times. This steep dependence on chain length is what amplifies differences in $\koff$. The Markov chain studied here (Sec.~\ref{sec:model}) has the same structure, with rebinding after each dissociation, and ligand quality enters it through $\koff$ measured in units of $\kp$.
 
One might equate $\nkp$ with the number of ITAM phosphorylation events, but the correspondence is loose. TCR signaling offers many candidate transitions, among them ITAM phosphorylation, ZAP70 binding and activation, LAT phosphorylation, signalosome assembly and downstream enzymatic reactions. Several ITAMs can be phosphorylated in parallel, individual ITAMs are partly redundant, and reactions downstream of the receptor add delays of their own. In a coarse-grained model, $\nkp$ is better read as the effective number of sequential discriminating steps between ligand binding and whichever readout is measured. Ten ITAMs do not imply $\nkp=10$, and an inferred $\nkp$ above 10 does not require more than ten ITAMs.
 
Experiments support this reading. Optogenetic systems in which ligand lifetime can be tuned independently of other binding parameters have shown that T cell signaling depends on binding duration as kinetic proofreading predicts~\cite{yousefi2019,tischer2019}. Using a light-gated chimeric antigen receptor, Britain, Town and Weiner measured proofreading strength at successive points in the cascade and found an effective $\nkp=4.5\pm0.4$ at ZAP70 recruitment, $7.8\pm1.1$ at LAT clustering and $11.3\pm1.5$ at DAG generation~\cite{britain2022}. Proofreading thus appears to be spread across the proximal signaling network, with each downstream checkpoint adding discrimination. The range $\nkp=2$--10 considered in this work covers the first two readouts and reaches the lower end of the third.
 
Binding, unbinding and every proofreading step are random events. For a single TCR/pMHC pair, the activation time is therefore a first-passage time, the first time a stochastic trajectory started in the bound state reaches the final proofreading state~\cite{li2024}. A T cell, however, engages many ligands at once. If it activates as soon as any one of $n$ approximately independent pairs completes the chain, the relevant random variable is the fastest of their first-passage times, $T_{1,n}=\min(T_1,\ldots,T_n)$~\cite{morgan2023}. Combining kinetic proofreading with this extreme statistic connects the kinetics of single receptors to the timing of cellular activation.
 
Morgan and Lindsay~\cite{morgan2023} showed that for large $n$ this minimum approaches a Weibull distribution. The argument uses only the short-time behavior of a single pair. Before dissociation and rebinding have had time to matter, a pair can activate only by making all $\nkp$ forward jumps in a row, so its activation probability grows as the $\nkp$th power of time, with a prefactor that depends on $\kp$ and $\nkp$. Extreme-value theory then turns this power law into a Weibull law for $T_{1,n}$ with shape parameter $\nkp$ and a scale that decreases with $n$ [Eq.~(1) in Sec.~\ref{sec:model}]. A matrix-exponential first-passage calculation is thereby replaced by a closed form whose dependence on receptor number and chain length can be read off directly. Morgan and Lindsay compute their results from the full system of ordinary differential equations and use the Weibull form to identify the large-$n$ limit, so the accuracies they report do not rest on it. It is, however, the closed form a reader would naturally reach for.
 
The prefactor of the short-time power law, and with it the Weibull law, contains $\kp$ and $\nkp$ but not $\koff$. Two ligands with the same $\kp$ and off-rates a hundredfold apart are therefore assigned the same activation-time distribution, although telling such ligands apart is the reason kinetic proofreading was invoked in the first place. As an asymptotic result the Weibull law is correct: for large enough $n$, the first activation happens so early that dissociation has no time to affect the successful trajectories. The trouble comes from using it outside that short-time window, where it silently discards the process that governs longer times.
 
Faster unbinding shortens this window, deeper chains push the activation time toward its edge, and more engaged receptors pull it back. Inside the window, the first receptor to activate has usually run through the whole chain during its initial binding event, so activation is a race among first attempts, and the Weibull law describes the winner's time well. Outside it, an uninterrupted pass through the chain is rare. Pairs bind, advance part of the way, lose their ligand and start over many times before one attempt succeeds, and the waiting time accumulated in these failed cycles dominates the first-passage time. Because the probability of an uninterrupted pass is the per-step success probability raised to the power $\nkp$, it becomes very small for short-lived ligands and long chains.
 
Parts of this picture are already known. For a single proofreading chain, Bel, Munsky and Nemenman found that as the chain lengthens, the completion time becomes either nearly deterministic or nearly exponential, with a narrow transition between the two~\cite{bel2010,munsky2009}; the collapse of the effective Weibull shape of $T_{1,n}$ that we report in Sec.~\ref{sec:shape} is the extreme-value counterpart of that dichotomy. Whether additional proofreading steps make discrimination more reliable once stochasticity is included has been debated~\cite{kirby2023}, with answers that change when the time available to decide grows with the number of steps~\cite{xiao2024} and that depend on whether the cell reads out a first-passage time or a count of signaling products~\cite{li2024}.
 
The competition between a fast and a slow regime is not peculiar to proofreading either. Lawley extended the theory of fastest first-passage times to random walks on networks, including a way to estimate whether a given system is in the fast regime~\cite{lawley2020networks}, and Madrid and Lawley derived conditions that place a diffusive search in a fast regime, set by the number of searchers, or a slow regime, set by single-searcher kinetics, with approximations of the full distribution in each~\cite{madrid2020}. They also showed that where the searchers start can matter a great deal. In the present chain, pairs that start bound give a Weibull shape of $\nkp$, and pairs that start unbound give $\nkp+1$. When each individual passage is a rare event, MacLaurin and Newby found a joint scaling of searcher number and rarity under which the mean fastest time stays finite~\cite{maclaurin2025}, and Grebenkov, Metzler and Oshanin related the fastest first-passage time of $N$ random walkers to macroscopic reaction rates~\cite{grebenkov2020}. Lawley reviews these results, including search by discrete jumps~\cite{lawley2024}. To our knowledge, two things have not been worked out explicitly for the proofreading chain: a test of which regime the fastest of $n$ receptors is in, written in terms of $\kp$, $\koff$, $\nkp$ and $n$, and a closed form that follows the activation time across the crossover without losing $\koff$.

Here we fill both gaps for the chain of Sec.~\ref{sec:model}. We treat each of $n$ independent TCR/pMHC pairs as a continuous-time Markov chain with $\nkp$ proofreading steps, forward rate $\kp$, dissociation rate $\koff$ and rebinding rate $k_1$, and compare the exact first-passage law with stochastic simulations and with the Weibull law. How far the Weibull law can be trusted is set by a single dimensionless number, $\Lambda$ [Eq.~\eqref{eq:lambda}], which measures the Weibull-predicted activation time in units of $\nkp+1$ mean bound-state lifetimes and factorizes into a part set by ligand quality and a part set by $\nkp$ and $n$. To leading order the relative error of the Weibull law equals $\Lambda$, and because $\Lambda$ is computed from the Weibull prediction itself, the test can be applied before any exact calculation. For $\Lambda\ll1$ the fastest pair activates during its first binding event and the Weibull law is accurate. As $\Lambda$ approaches one, dissociation and rebinding take over, and the Weibull prediction can miss the true activation time by orders of magnitude.
 
The change of regime alters the shape of the activation-time distribution as well as its scale (Sec.~\ref{sec:shape}). With an effective Weibull shape near $\nkp$, activation behaves like a fairly sharp timer: contacts well short of the typical activation time rarely lead to signaling, while contacts somewhat longer almost always do. When retries dominate, the effective shape falls toward one and activation becomes nearly exponential, so the chance of activating in the next moment no longer depends on how long the contact has already lasted. Proofreading still suppresses poor ligands in this limit, because successful passes are rare, but the timing precision of a multistep chain is lost. On our parameter grid, faster unbinding and longer chains both move activation from timer-like toward memoryless behavior.
 
To restore $\koff$, we split the first-passage time of a single pair into the duration of its eventual successful pass through the chain and the time spent in the failed binding cycles before it, and replace the geometric sum of failed cycles by its exponential limit (R\'enyi's theorem~\cite{kalashnikov1997}). The resulting closed form has two terms, one for the first attempt and one for the stream of retries behind it [Eq.~\eqref{eq:F1lin}]. It reduces to the Weibull law when activation precedes any dissociation, becomes an exponential first-activation law with an explicit trigger rate [Eq.~(8)] when retries dominate, and reproduces the collapse of the effective shape between these limits. On the $5\times5$ grid studied here, its median lies within 2\% of the exact median in 19 of the 25 cells and within about 6\% in all of them, whereas the Weibull median is off by a factor of up to 2900.

What substituting the Weibull form would cost is clearest in the agonist-versus-self classifier of Ref.~\cite{morgan2023}. Because the Weibull law contains no $\koff$, it treats agonist and self ligands as identical, so an APC with one agonist among $10^4$ self pMHCs would differ from one with none only by a single extra ligand, and discrimination would fall to chance. For ten proofreading steps and the fastest-unbinding ligands on our grid, the Weibull law also predicts that self pMHCs trigger activation in under two seconds, whereas the exact law and our closed form give roughly an hour at the rebinding rate used here.

\section{Model}\label{sec:model}

Consider one TCR/pMHC signaling pair as a continuous-time Markov chain on $\{U,C_0,\ldots,C_{\nkp}\}$ with $C_{\nkp}$ absorbing:
\[
U\ \xrightarrow{\ k_1\ }\ C_0\ \xrightarrow{\ \kp\ }\ C_1\ \xrightarrow{\ \kp\ }
\ \cdots\ \xrightarrow{\ \kp\ }\ C_{\nkp}, \qquad C_i\ \xrightarrow{\ \koff\ }\ U\quad (i=0,\ldots,\nkp-1),
\]

following McKeithan's proposal that a chain of forward modification steps, each undone by dissociation, lets a receptor discriminate on $\koff$ rather than on affinity~\cite{mckeithan1995}. Each pair is initialized in $C_0$ at $t=0$; the rate $k_1$ enters only by rebinding after a dissociation. Pairs that start unbound instead need one more jump to reach $C_{\nkp}$, which raises the Weibull shape to $\nkp+1$. Assuming the $n$ pairs are independent, \[
P(T_{1,n}\le t)\;=\;1-S(t)^{n}
\;=\;1-\Bigl[\,1-\bigl(\mathrm{e}^{Qt}\bigr)_{C_0,C_{\nkp}}\Bigr]^{n} .
\tag{exact}
\]
Here $Q$ is the generator of the chain, $S(t)$ is the probability that a single pair has not activated by time $t$, and $T_{1,n}$ is the first activation time among the $n$ pairs. Independence is an idealization: in Ref.~\cite{morgan2023} the pairs interact through receptor and ligand binding, and the independent-pair law is the reduction analyzed here. Eq.~(1) of~\cite{morgan2023} is the $n\to\infty$ limit of this, with
$\lambda_n=(An)^{-1/\nkp}$ and $A=\kp^{\nkp}/\nkp!$,
\begin{equation}
P\bigl(T_{1,n}\le \lambda_n x\bigr)\ \longrightarrow\ 1-\mathrm{e}^{-x^{\nkp}},
\qquad n\to\infty,
\tag{1}
\end{equation}
that is, $T_{1,n}/\lambda_n \Rightarrow \mathrm{Weibull}(1,\nkp)$, used in
practice in the form $T_{1,n}\approx\mathrm{Weibull}(\lambda_n,\nkp)$.

The closed form is missing $\koff$, the unbinding rate. This means that two ligands whose off-rates differ by a hundredfold are assigned the same predicted activation time. 

We measure time in units of $1/\kp$, equivalently setting
$\kp=1~\mathrm{s}^{-1}$, so, every time reported in Table~\ref{tab:medians} and in Figs.~\ref{fig:eq1grid}
and~\ref{fig:pdfgrid} is in these units. Beyond Eq.~(1)'s own variables the parameter space is then two ratios,
$\om=\koff/\kp$ and $k_1/\kp$. The grid is run at $k_1/\kp = 1$. Throughout, the relative error of Eq.~(1) at a
given quantile is
\[
E \;=\; 1-\frac{t_{\mathrm{Eq}(1)}}{t_{\mathrm{exact}}},
\]
so that $E\to1$ when Eq.~(1) underestimates the activation time without
bound. Table~\ref{tab:knobs} gives the effect of each remaining parameter on $E$.

\begin{table}[htbp]
\centering
\small
\caption{Every remaining parameter, and what it does to the relative error of
the Eq.~(1) median at $n=10^4$. Only $\om$ has a large effect that is
\emph{not} entangled with $\nkp$. Relative error is $E$ as defined in Sec.~\ref{sec:model} and $t_q$ is the Eq.~(1) median, in units of $1/\kp$. The $k_1/\kp$ row is evaluated at $\nkp=3$, $\om=1$ and $n=10^4$, inside the Weibull regime; once retries dominate, $k_1$ enters the activation time at first order through Eq.~(8) (Sec.~\ref{sec:biologicalsignificance}).}
\label{tab:knobs}
\begin{tabularx}{\textwidth}{|X|X|c|}
\toprule
variable & effect on the Eq.~(1) median error & coupled to $\nkp$? \\
\midrule
$\om=\koff/\kp$ & multiplies it by $(1+\om)/2$ relative to the $\om=1$ column,
to leading order in $\Lambda$ & no \\
$\nkp$ & grows it like $t_q/(\nkp+1)$, $t_q=(\nkp!\ln 2/n)^{1/\nkp}$ & n/a  \\
$n$ (ligand/receptor number) & shrinks it like $n^{-1/\nkp}$ &
  yes  \\
$k_1/\kp$ (rebinding) & $10^{-2}\to10^{4}$ moves it $0.0379\to0.0314$ &
  second order 
  \\
all rates $\times c$ & nothing: the error is dimensionless & n/a \\
\bottomrule
\end{tabularx}
\end{table}

\section{Validity of the Weibull limit}\label{sec:validity}

\begin{figure}[htbp]
\centering
\includegraphics[width=\textwidth]{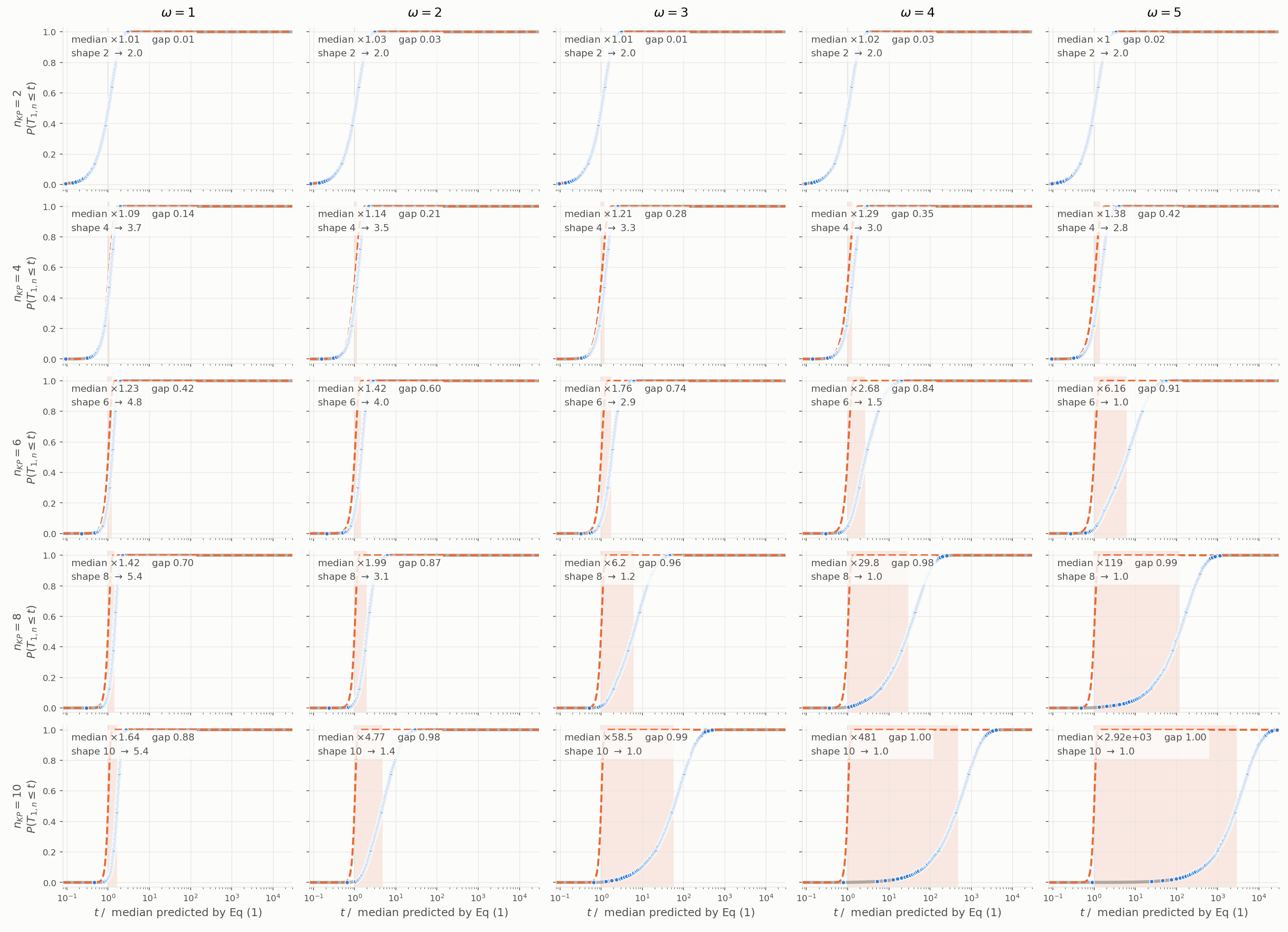}
\caption{Simulated $T_{1,n}$ (dots), the exact law (grey), and Eq.~(1) (dashed),
on a shared axis of $t/\text{median}_{\mathrm{Eq}(1)}$. 
Annotations per cell: ``median $\times$'' is the simulated median divided by
the Eq.~(1) median; ``gap'' is the largest vertical distance between the
simulated and Eq.~(1) curves; ``shape $a\to b$'' gives the shape Eq.~(1) asserts
($a=\nkp$) against the effective shape of the exact law ($b$,
Table~\ref{tab:shape}).
Shaded band in each cell spans the displacement of the median from Eq.~(1)'s prediction to the simulation. With 6000 replicates, the Monte
Carlo standard error on the median is
$0.019/k$ for effective shape $k$, about $0.9\%$ at $k=2$, rising to about
$1.9\%$ in the cells where the shape has collapsed to $1$.}
\label{fig:eq1grid}
\end{figure}

Figure~\ref{fig:eq1grid} puts the simulation and Eq.~(1) side by side over a
$5\times5$ grid. Rows add proofreading steps and columns raise $\om$. Because
$\om$ does not appear in Eq.~(1), \emph{all five cells of a row are measured
against the identical dashed prediction}, so any variation along a row is error.

The top row barely moves because at $\nkp=2$ activation
completes inside one
bound-state lifetime and $\om$ shifts the median by only $1.6\%$ across the whole row, which the Monte Carlo scatter in Table~\ref{tab:medians} obscures. But moving down and to the right the dashed curve starts to detach, reaching a simulated median $2.9\times10^{3}$
times Eq.~(1)'s in the bottom-right cell. 

\begin{table}[htbp]
\centering\small
\caption{Median activation time (s), simulated, $n=10^{4}$, 6000 contacts per
cell, at $\kp=1~\mathrm{s}^{-1}$. To read these against
Table~\ref{tab:bridge}, divide by $\nkp/8$.}
\label{tab:medians}
\begin{tabular}{@{}lrrrrr@{}}
\toprule
& $\om=1$ & $\om=2$ & $\om=3$ & $\om=4$ & $\om=5$ \\
\midrule
$\nkp=2$  & 0.0119 & 0.0121 & 0.0118 & 0.0120 & 0.0118 \\
$\nkp=4$  & 0.221 & 0.230 & 0.243 & 0.260 & 0.278 \\
$\nkp=6$  & 0.749 & 0.863 & 1.07 & 1.63 & 3.74 \\
$\nkp=8$  & 1.61 & 2.27 & 7.05 & 33.9 & 135 \\
$\nkp=10$ & 2.86 & 8.29 & 102 & 835 & $5.08\times10^{3}$ \\
\bottomrule
\end{tabular}
\end{table}

\begin{figure}[htbp]
\centering
\includegraphics[width=\textwidth]{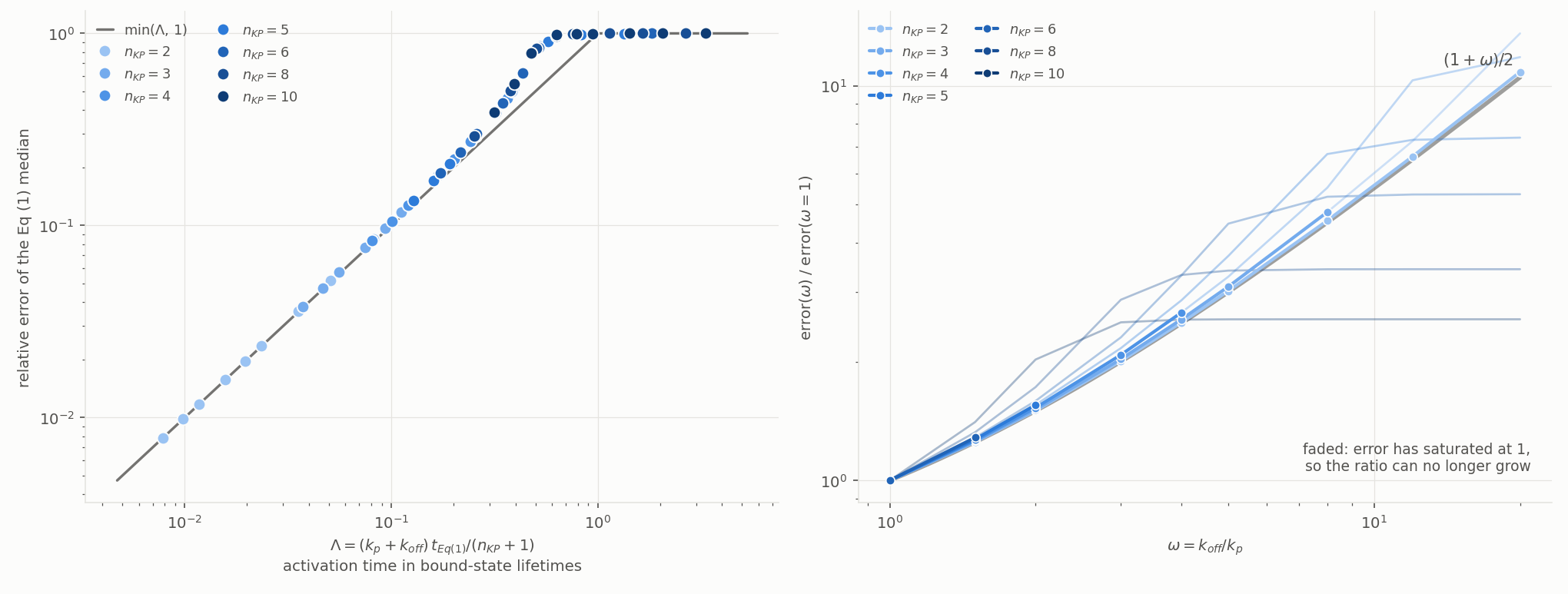}
\caption{Left: all 63 $(\nkp,\om)$ pairs of the extended sweep
$\nkp\in\{2,3,4,5,6,8,10\}$, $\om\in\{1,1.5,2,3,4,5,8,12,20\}$ collapse onto
$\mathrm{error}=\min(\Lambda,1)$. The measured error \emph{exceeds}
$\Lambda$ by a relative amount that itself grows like $\Lambda$, from under $2\%$ for
$\Lambda<0.05$ to roughly $24\%$ at $\Lambda=0.32$. Right: dividing out the $\om=1$ column leaves
$(1+\om)/2$ at every $\nkp$. Faded segments are cells whose error has
saturated at $1$, a ceiling on the \emph{measure}.}
\label{fig:sep}
\end{figure}

The single-pair law behind Eq.~(1) is only the first term of a series,
\[
F(t)=A\,t^{\nkp}\left[1-\frac{\nkp(\kp+\koff)}{\nkp+1}\,t+O(t^{2})\right],
\]
so the relative error of the Eq.~(1) quantile is
\begin{equation}
\Lambda \;=\; \frac{(\kp+\koff)\,t_q}{\nkp+1}
\;=\;\underbrace{(1+\om)}_{\om\text{ only}}\times
\underbrace{\frac{t_q}{\nkp+1}}_{\nkp,\,n\text{ only}} ,
\label{eq:lambda} \tag{2}
\end{equation}
with $t_q$ in units of $1/\kp$ as above.
Equivalently, $F(t)/(A t^{\nkp})$ depends on $\om$ \emph{only} through the
product $(\kp+\koff)t$. Across $\omega$ from 1 to 200, at fixed $(k_{\rm p}+k_{\rm off})t\lesssim1$ and fixed
$k_1/k_{\rm p}$, that ratio moves by at most 2.1\% (at $n_{\rm KP}=2$), and by under 0.5\% for $n_{\rm KP}\ge6$.
Only past $(k_{\rm p}+k_{\rm off})t\approx2$ does the residual $\omega$ dependence reach
several percent. So $\om$ does not change the shape of
Eq.~(1)'s validity window; it shrinks the window by a factor $(1+\om)$, and
$\nkp$ and $n$ decide where the activation time sits inside it. Figure~\ref{fig:sep} shows both halves: the error at all 63 points of an extended sweep collapses onto $\min(\Lambda,1)$, and dividing out the $\om=1$ column leaves $(1+\om)/2$ at every $\nkp$.

The two parameters enter the error of Eq.~(1) as separate factors, but their effects on the activation time itself interact strongly: raising $\om$ fivefold barely moves the median of a two-step chain but lengthens that of a ten-step chain more than a thousandfold (Table~\ref{tab:medians}; Sec.~\ref{sec:biologicalsignificance}).

\section{Effect on the shape of the activation-time distribution}\label{sec:shape}

\begin{figure}[htbp]
\centering
\includegraphics[width=\textwidth]{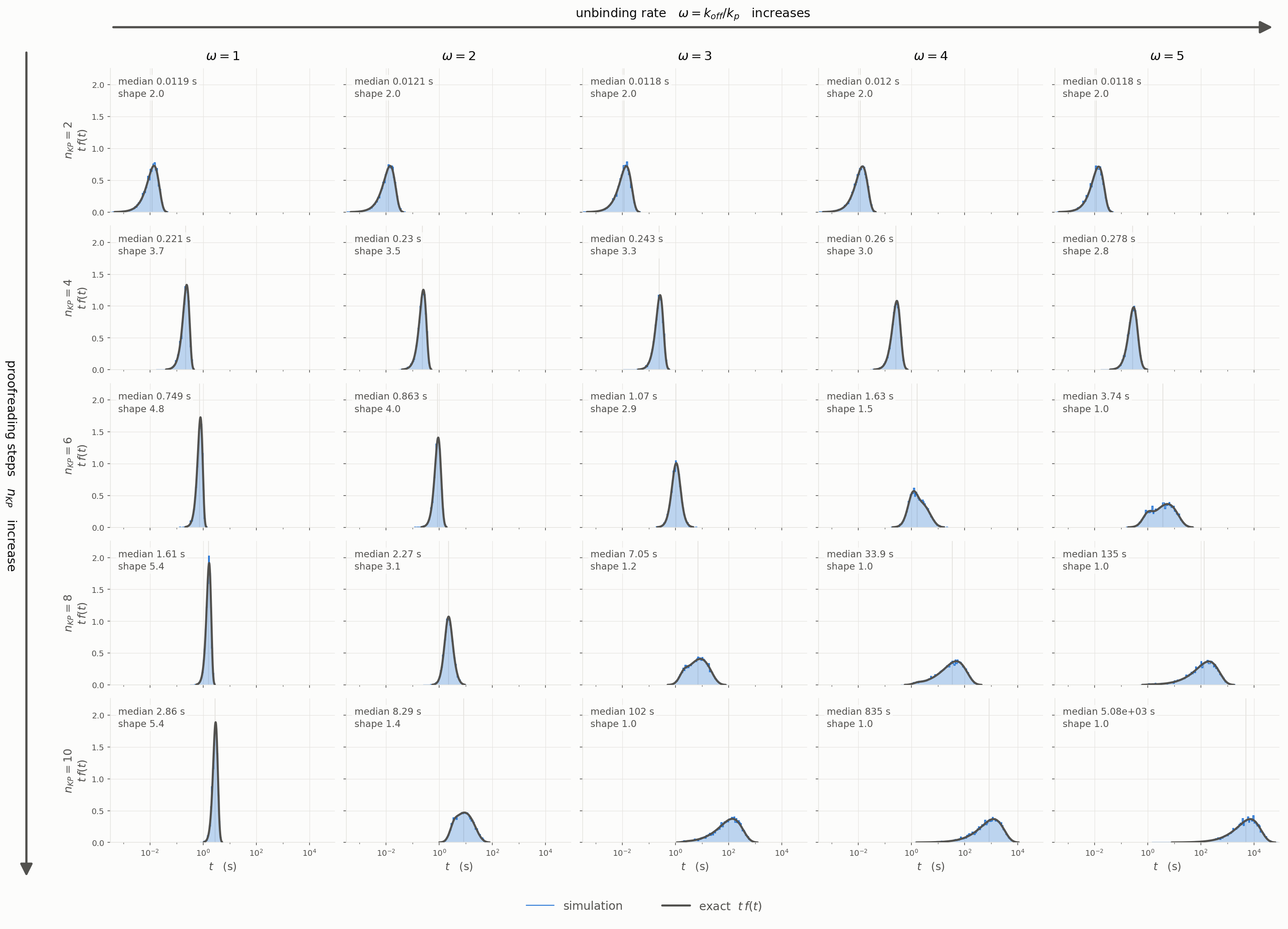}
\caption{Densities of $T_{1,n}$ on one shared $t$ axis and one shared
$t\,f(t)$ axis. Blue: 6000 simulated contacts per cell, bins a fixed $0.04$
decades wide. Black: the exact $t\,f(t)$. Down a column the peak grows taller
and narrower. Across a row it collapses.}
\label{fig:pdfgrid}
\end{figure}

Figure~\ref{fig:pdfgrid} shows the distributions when $t\,f(t)$ is plotted against $\log t$. This is done because the
integral of $t f(t)$ over $\mathrm{d}\log t$ is $1$ for every parameter set.

\begin{table}[htbp]
\centering\small
\caption{Effective Weibull shape, measured as the slope of $\log(-\log S)$ over
the interquartile range of the exact law. Eq.~(1) asserts every entry equals its
row label.}
\label{tab:shape}
\begin{tabular}{@{}lrrrrr@{}}
\toprule
& $\om=1$ & $\om=2$ & $\om=3$ & $\om=4$ & $\om=5$ \\
\midrule
$\nkp=2$  & 2.0 & 2.0 & 2.0 & 2.0 & 2.0 \\
$\nkp=4$  & 3.7 & 3.5 & 3.3 & 3.0 & 2.8 \\
$\nkp=6$  & 4.8 & 4.0 & 2.9 & 1.5 & 1.0 \\
$\nkp=8$  & 5.4 & 3.1 & 1.2 & 1.0 & 1.0 \\
$\nkp=10$ & 5.4 & 1.4 & 1.0 & 1.0 & 1.0 \\
\bottomrule
\end{tabular}
\end{table}

An effective shape near $\nkp$ means that activation behaves like a \emph{timer}: contacts well short of the typical activation time rarely lead to signaling, while contacts somewhat longer almost always do. An effective shape of $1$ means the activation time is exponential and therefore memoryless: there is no safe contact duration, and the probability of a false positive grows steadily with contact length. Table~\ref{tab:shape} shows that for $\nkp\ge6$ the effective shape falls well short of $\nkp$ even at $\om=1$, and that it reaches $1$ in the right-hand columns.

\section{A closed form that retains $\om$}\label{sec:closed}
Equation~(1) discards $\omega$ because it keeps only the leading term of the
single-pair law. Restoring it does not require abandoning a closed form.
Write $\mu = k_{\rm p}+k_{\rm off} = k_{\rm p}(1+\omega)$ for the escape rate from a
bound state and $r = k_{\rm p}/\mu = (1+\omega)^{-1}$ for the probability that a given
escape is forward rather than a dissociation. An attempt beginning in $C_0$ reaches
$C_{n_{\rm KP}}$ with probability
\[
\sigma = r^{\,n_{\rm KP}} = (1+\omega)^{-n_{\rm KP}},
\]
and, conditional on success, takes exactly $\mathrm{Gamma}(n_{\rm KP},\mu)$. The pair
makes $n_{\rm KP}$ jumps with i.i.d.\ $\mathrm{Exp}(\mu)$ holding times. The first
passage time of a single pair therefore splits exactly as
\[
T \;=\; W + G, \qquad G \sim \mathrm{Gamma}(n_{\rm KP},\mu), \qquad W \perp G,
\]
where $W=\sum_{i=1}^{M-1}X_i$ is the time consumed by failed attempts,
$M\sim\mathrm{Geom}(\sigma)$, and the $X_i$ are i.i.d.\ failed cycles. Their mean is
\[
\tau_{\rm c} \;=\; \frac{1}{k_1}
+ \frac{1}{1-\sigma}\left(\frac{1-\sigma}{k_{\rm off}}-\frac{n_{\rm KP}\sigma}{\mu}\right)
\;\xrightarrow[\;\sigma\ll 1\;]{}\; \frac{1}{k_1}+\frac{1}{k_{\rm off}} .
\]

The single assumption we add is that the geometric
sum $W$ is exponential apart from its atom at the origin,
\begin{equation}
W \approx \sigma\,\delta_0 + (1-\sigma)\,\mathrm{Exp}(\lambda),\qquad
\lambda=\frac{\sigma}{\tau_{\rm c}}\;\;(<\mu),
\label{eq:assumption}
\tag{3}
\end{equation}
which matches $\mathbb{E}[W]$ exactly and is the standard small-$\sigma$ limit of a
geometric compound~\cite{kalashnikov1997}. The atom carries the pairs that trigger without ever dissociating
and is what preserves the short-time behavior Eq.~(1) is built on. Convolving
Eq.~(\ref{eq:assumption}) with $G$ gives a closed form for one pair,
\begin{equation}
F_1(t) \;=\; \Pinc(n_{\rm KP},\mu t)
\;-\;(1-\sigma)\,e^{-\lambda t}\!\left(1-\tfrac{\lambda}{\mu}\right)^{-n_{\rm KP}}
\Pinc\!\big(n_{\rm KP},(\mu-\lambda)t\big),
\label{eq:F1}
\tag{4}
\end{equation}
with $\Pinc(a,x)=1-e^{-x}\sum_{j<a}x^{j}/j!$ the regularized lower incomplete
gamma function. Because $\lambda t\ll1$ everywhere the minimum of $n$ pairs sits,
Eq.~(\ref{eq:F1}) is equivalent, to the order kept, to the more transparent
\begin{equation}
F_1(t) \;=\; \sigma\left[\,\Pinc(n_{\rm KP},\mu t)
\;+\;\frac{1}{\tau_{\rm c}}\left(t\,\Pinc(n_{\rm KP},\mu t)
-\frac{n_{\rm KP}}{\mu}\,\Pinc(n_{\rm KP}+1,\mu t)\right)\right],
\label{eq:F1lin}
\tag{5}
\end{equation}
whose two terms are the first attempt and the steady stream of retries behind it. Eq.~(5) is a small-$F_1$ truncation as it grows without bound in $t$ and is to be used
where $nF_1=O(1)$, which is where the minimum of $n\gg1$ pairs sits. Past that, we clip it at unity or evaluate Eq.~(4). It also drops a factor $1-\sigma$ from the retry term,
which moves no median on the grid by more than $10^{-3}$. For
$n$ independent pairs the extreme statistic is exact given $F_1$,
\begin{equation}
P(T_{1,n}\le t) \;=\; 1-\big(1-F_1(t)\big)^{n},
\tag{6}
\end{equation}
and the $q$ quantile solves $F_1(t_q)=1-(1-q)^{1/n}$, a monotone one-dimensional root
find. Writing $L=-\ln(1-q)$, the quantiles of the two limiting laws are
\begin{equation}
t_q^{\rm W}=\left(\frac{n_{\rm KP}!\,L}{n\,k_{\rm p}^{n_{\rm KP}}}\right)^{1/n_{\rm KP}},
\qquad
t_q^{\rm E}=\frac{n_{\rm KP}}{\mu}+\tau_{\rm c}\left(\frac{L}{n\sigma}-1\right),
\label{eq:brackets}
\tag{7}
\end{equation}
and $t_q^{\rm W}$ is exactly the Eq.~(1) quantile. 

The first is a lower bound and the second is not. Completing the chain requires $n_{\rm KP}$ forward jumps, which occur only while the pair is bound and therefore no faster than the ticks of a rate-$k_{\rm p}$ Poisson clock, so $F(t)\le\Pinc(n_{\rm KP},k_{\rm p}t)\le At^{n_{\rm KP}}$. Hence $t_q^{\rm W}$ lies below the exact quantile, up to a relative correction of order $L/(n\,n_{\rm KP})$, and in every cell of the grid it also lies below the root of Eq.~(5). By contrast, $t_q^{\rm E}$ is negative in nine of the 25 cells. An upper bound follows from writing the retry term of Eq.~(5) as $\sigma\,\mathbb{E}[(t-G)^+]/\tau_{\rm c}$ and applying Jensen's inequality, $\mathbb{E}[(t-G)^+]\ge t-n_{\rm KP}/\mu$, which gives $F_1(t)\ge(\sigma/\tau_{\rm c})(t-n_{\rm KP}/\mu)$ and hence
\[
t_q\;\le\;t_q^{\rm U}\;=\;\frac{n_{\rm KP}}{\mu}+\frac{\tau_{\rm c}}{\sigma}\Bigl[1-(1-q)^{1/n}\Bigr].
\]
The root of Eq.~(5) therefore lies in $[t_q^{\rm W},t_q^{\rm U}]$, from which bisection or Brent's method converges.

When $\mu t_q\ll1$ the retry term is negligible and $F_1\to A t^{n_{\rm KP}}$, so
Eq.~(5) collapses onto Eq.~(1), and the cost of that collapse is the $\Lambda$ of Eq.~(2). When the
first attempt is hopeless, $F_1\to\kappa t$ with an effective trigger rate
\begin{equation}
\kappa \;=\; \frac{\sigma}{\tau_{\rm c}}
\;=\; \frac{(1+\omega)^{-n_{\rm KP}}}{k_1^{-1}+k_{\rm off}^{-1}}\quad(\sigma\ll1),
\tag{8}
\end{equation}
so $T_{1,n}\to\mathrm{Exp}(n\kappa)$ and the effective shape is $1$. The collapse
tabulated in Sec.~\ref{sec:shape} is therefore not a separate phenomenon; it is the retry term of
Eq.~(\ref{eq:F1lin}) overtaking the first-attempt term, and $\omega$ drives it because
$\sigma$ carries $\omega$ to the $n_{\rm KP}$th power.

\begin{figure*}[t]
\includegraphics[width=\textwidth]{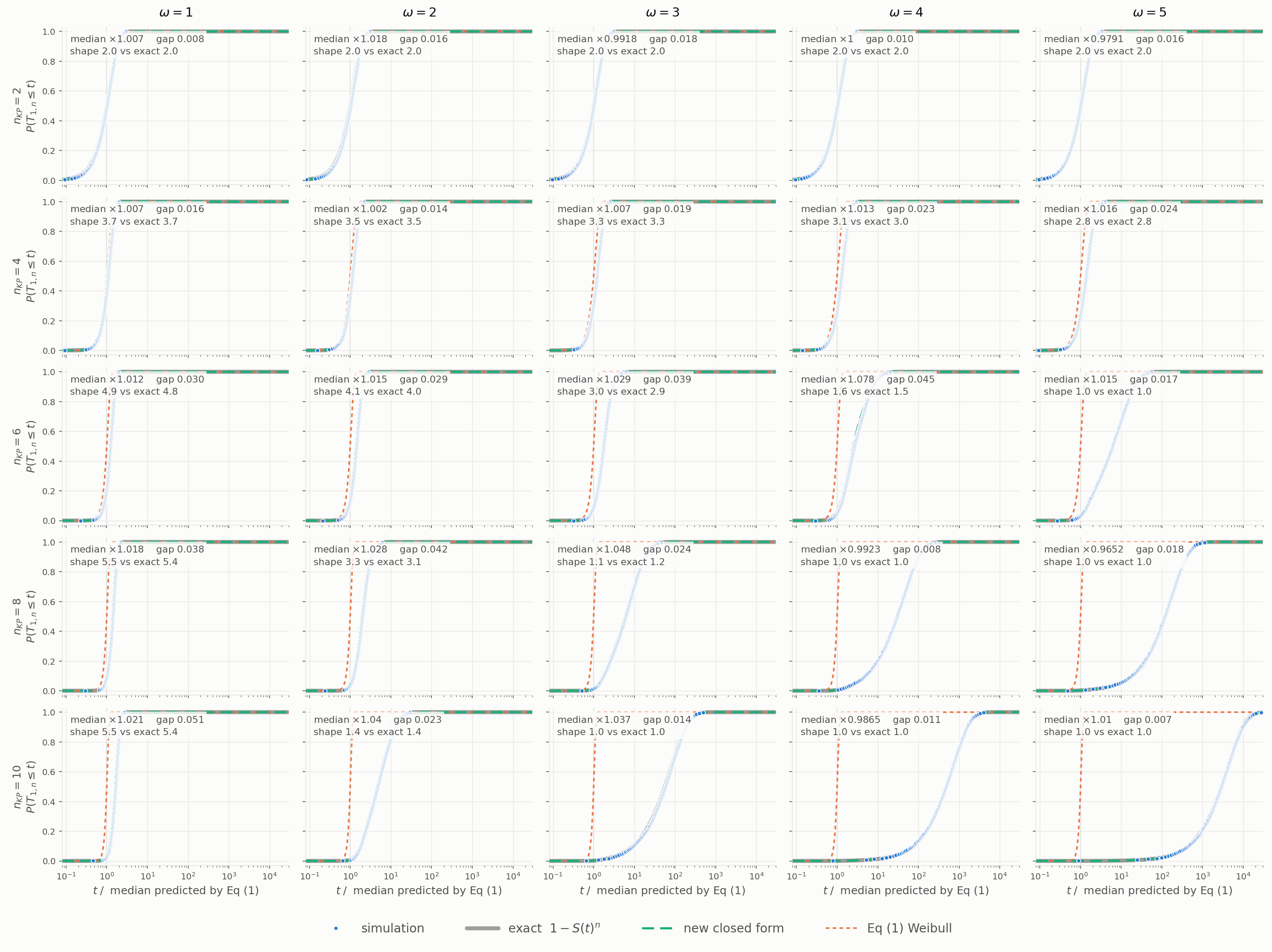}
\caption{Eq.~(5) against simulation and the exact law, on the axes of Fig.~\ref{fig:eq1grid}. Dots: the same 6000 simulated contacts per cell as in Fig.~\ref{fig:eq1grid}; Grey: the exact law $1-S(t)^n$; Green dashed: the
closed form, Eq.~(5) substituted into Eq.~(6), hidden beneath the gray curve wherever
the two agree; Orange dashed: Eq.~(1), the same curve in all five cells of a row;
Annotations per cell: ``median $\times$'' is the simulated median divided by
Eq.~(5)'s median, the same ratio Fig.~\ref{fig:eq1grid} reports against Eq.~(1), ``gap'' is the
largest vertical distance between the simulated and Eq.~(5) curves, ``shape'' is the
effective Weibull shape of Eq.~(5) against that of the exact law (Table~\ref{tab:shape}), measured
by the same inter-quartile slope. Monte Carlo standard error on the median is as in
Fig.~\ref{fig:eq1grid}.}
\label{fig:newgrid}
\end{figure*}

Figure~\ref{fig:newgrid} places Eq.~(5) on the axes of Fig.~\ref{fig:eq1grid}. The simulated median falls within 5\% of
Eq.~(5)'s in 24 of the 25 cells and within 3\% in 20, against a Monte Carlo standard
error of 0.9--1.9\% on the median itself. The largest vertical distance between the simulated and Eq.~(5) curves is 0.051,
against 0.019 for the simulator against the exact law (Table~\ref{tab:r2}), so what remains in
those cells is the closed form's residual rather than the sampler's. Measured against
the exact law directly, that distance peaks at 0.044 at $n_{\rm KP}=8$, $\omega=2$ and
exceeds 0.019 in ten of the 25 cells, all of them cells where the retry term carries
between 6\% and 78\% of $F_1$ at the median. Wherever one mechanism carries more than
95\%, the distance never exceeds 0.019. Eq.~(5) also reproduces
the shape collapse of Table~\ref{tab:shape} cell by cell, including the crossover row
$n_{\rm KP}=6$, where the effective shape falls from 4.9 to 1.0 across the five values
of $\omega$ while Eq.~(1) asserts 6 throughout.

\begin{table}[htbp]
\centering\small
\caption{Relative error of the Eq.~(5) median against the exact law, in percent,
$n=10^{4}$, $k_1/\kp=1$. A positive value indicates that the Eq.~(5) median lies below the
exact median. The same measure for Eq.~(1) runs from $0.78$ at
$\nkp=2,\ \om=1$ to $99.97$ at $\nkp=10,\ \om=5$.}
\label{tab:eq5err}
\begin{tabular}{@{}lrrrrr@{}}
\toprule
& $\om=1$ & $\om=2$ & $\om=3$ & $\om=4$ & $\om=5$ \\
\midrule
$\nkp=2$  & 0.12 & 0.14 & 0.15 & 0.16 & 0.17 \\
$\nkp=4$  & 0.62 & 0.85 & 1.05 & 1.24 & 1.47 \\
$\nkp=6$  & 1.06 & 1.78 & 2.96 & 6.12 & 4.33 \\
$\nkp=8$  & 1.46 & 3.58 & 3.59 & 0.58 & 0.12 \\
$\nkp=10$ & 1.92 & 4.01 & 0.26 & 0.03 & 0.01 \\
\bottomrule
\end{tabular}
\end{table}

The quantile is no longer explicit, though it costs only a one-dimensional root find. The memory-less retry
stream in Eq.~(\ref{eq:assumption}) can fire before a cycle has physically elapsed,
which adds a spurious term of relative size $t/[(n_{\rm KP}+1)\tau_{\rm c}]$ to the
short-time expansion. It is smaller than the genuine $O(\Lambda)$ term of the same expansion by a factor
$\nkp\mu\tau_{\rm c}$, between 6.7 and 72 across the grid. Both approximations are mildest in the two limits and worst between them where the residual sits. Measured against the exact law rather than against the
simulation (so that Monte Carlo scatter is removed), Eq.~(5)'s median is within 2\% in 19 of the 25 cells (Table~\ref{tab:eq5err}) and its single largest error is 6\% at $n_{\rm KP}=6$, $\omega=4$,
the cell in which the first attempt and the retry stream contribute comparably.

The $5\times5$ grid holds $k_1/\kp=1$ and $n=10^4$ fixed. Over a wider sweep of 1080 parameter sets, with $\om$ up to 28, $k_1/\kp$ spanning four decades and $n$ from $10^2$ to $10^5$, the worst error of Eq.~(5) at the 5th, 25th, 50th, 75th or 95th percentile is 12.5\%, reached at $n=10^2$, where the extreme-value target $-\ln(1-q)/n$ is largest; the relative error of Eq.~(1) reaches 1 in the same sweep.

\section{Biological significance}\label{sec:biologicalsignificance}

$\om=\koff/\kp$ is the ligand's unbinding rate in units of the proofreading
rate. It stands for antigen quality, the axis on which T cells discriminate.
Matsui et al.\ showed that the TCR/pMHC dissociation rate correlates with T cell
responsiveness, making the off-rate rather than the affinity the relevant measure
of quality~\cite{matsui1994}. Aleksic et al.\ measured the binding kinetics of a panel of pMHC variants for a single TCR at $37^{\circ}$C and found that both the on-rate and the off-rate set potency. They argued that the functionally relevant quantity is a \emph{confinement time} that includes rebinding, so the effective off-rate seen by the receptor is slower than the solution value~\cite{aleksic2010}.

Pettmann et al.\ then measured how much the response depends on the quality of the antigen.
Over a panel of pMHC reaching $K_D\sim1$ mM they measured discriminatory 
power $\alpha$, defined by potency $\propto K_D^{-\alpha}$, with a pooled 95\% CI of 1.9--2.1. At fixed on-rate this is equivalent to a $\koff^{-\alpha}$ dependence. Fitting a proofreading model to the same data, they found about 2.7 steps and a proofreading delay of 2.8~s~\cite{pettmann2021}.

Table~\ref{tab:bridge} maps $\om$ onto measured ligand half-lives. It uses
$\kp=\nkp/\tau_{\mathrm{KPR}}$ with $\tau_{\mathrm{KPR}}=8$~s, the threshold
binding half life measured optogenetically by Yousefi et al. (best fit
$8$~s, 95\% CI $3$--$19$~s)~\cite{yousefi2019}. Tischer and Weiner
reached the same conclusion by an independent optogenetic route~\cite{tischer2019}. Note that $\kp$ is
$\nkp$-dependent under this mapping, whereas the grid is run at
$\kp=1~\mathrm{s}^{-1}$. The dimensionless results are unaffected and absolute times
in Table~\ref{tab:medians} should be divided by $\nkp/8$ to be read against
Table~\ref{tab:bridge}.

\begin{table}[htbp]
\centering\small
\caption{$\om=\koff/\kp$ for ligands of a given binding half life, assuming
$\kp=\nkp/\tau_{\mathrm{KPR}}$ with $\tau_{\mathrm{KPR}}=8$~s~\cite{yousefi2019}.}
\label{tab:bridge}
\begin{tabular}{@{}lrrrrrr@{}}
\toprule
ligand & $t_{1/2}$ (s) & $\koff$ (s$^{-1}$)
& $\om$ at $\nkp{=}2$ & $\nkp{=}3$ & $\nkp{=}6$ & $\nkp{=}10$ \\
\midrule
strong agonist & 10.0 & 0.069 & 0.28 & 0.18 & 0.09 & 0.06 \\
weak agonist   & 2.0  & 0.347 & 1.39 & 0.92 & 0.46 & 0.28 \\
antagonist     & 0.7  & 0.990 & 3.96 & 2.64 & 1.32 & 0.79 \\
null / self    & 0.1  & 6.931 & 27.7 & 18.5 & 9.24 & 5.55 \\
\bottomrule
\end{tabular}
\end{table}

Table~\ref{tab:bridge} places the grid's column range, $\om=1$--$5$, between
weak agonist and antagonist, with genuine self ligands at or beyond its right
edge ($\om=5.6$--$28$). The effects we show are therefore at the mild end of the
physiological range.

$\nkp$ is a property of the receptor's signaling cascade as opposed to that of the encounter. It is fixed for a given receptor and slow to change, and its measured value depends on where in the pathway it is read out. With a light-gated chimeric antigen receptor, Britain, Town and Weiner measured proofreading strength at three successive nodes and obtained $\nkp=4.5\pm0.4$ at ZAP70 recruitment, $7.8\pm1.1$ at LAT clustering, and
$11.3\pm1.5$ at DAG generation~\cite{britain2022}. The range $\nkp=2$--$10$ spanned by our grid therefore brackets the first two readouts
and reaches the lower end of the DAG estimate.

The two axes previously explored have different biological significances. $\nkp$ is intrinsic to the cell and held constant during an encounter. $\om$ varies across the ligands present in a single contact by orders of magnitude. Because Eq.~\eqref{eq:lambda} factorizes, their effects on the formula's error multiply.

A receptor with more proofreading steps is \emph{more} sensitive to ligand quality. A fivefold change in off rate costs a two-step receptor next to nothing (exact median ratio 1.016) and costs a ten-step
receptor a factor of 1760. More proofreading buys specificity and
simultaneously buys a far longer wait.

The classifier of Ref.~\cite{morgan2023} gives an APC $n_{\mathrm{self}}=10^4$ self pMHC and $n_{\mathrm{ag}}=1$ agonist (a setup that assumes that a single agonist pMHC suffices to elicit a
calcium response, with $\sim\!10$ needed for synapse
formation~\cite{irvine2002}, and that a small number of complexes can trigger
killing~\cite{sykulev1996,purbhoo2004}).

Reference~\cite{morgan2023} evaluates the extreme statistic with a system of
ordinary differential equations and invokes Eq.~(1) only to identify the
$n\to\infty$ limit, so the accuracy reported there is unaffected by what follows.
Eq.~(1) is nonetheless the closed form a reader would reach for, and it is worth
recording what that substitution costs. If Eq.~(1) is applied to both populations, $\om$ becomes absent and the only difference between an agonist-positive and an agonist-negative APC is one extra
ligand out of ten thousand. Additionally, Eq.~(1)'s median scales as $n^{-1/\nkp}$, so
the \emph{self} population is predicted to fire first by $21.5\times$ at
$\nkp=3$ and $2.5\times$ at $\nkp=10$. With equal numbers of agonist-positive and agonist-negative APCs, the accuracy $\Gamma$ of Ref.~\cite{morgan2023} is the mean of two probabilities: that the T cell activates when an agonist is present, and that it stays silent when the agonist is absent. It then satisfies, for every $n_{\rm KP}$ and every $\omega$,
\[
\max_\tau \Gamma(\tau)=\frac12\left[1+\frac{1}{n+1}\left(\frac{n}{n+1}\right)^{n}\right]=0.500018
\quad\text{against a baseline of }0.5.
\]
For comparison, the exact independent-pair law reaches $\max_\tau\Gamma=0.9987$, at $\tau^{*}=1.7\times10^{4}$~s, for the parameters of Fig.~2B of Ref.~\cite{morgan2023}: $\nkp=10$, $\kp=1~\mathrm{s}^{-1}$, agonist and self off-rates of $1$ and $15~\mathrm{s}^{-1}$, and rebinding at $1~\mathrm{s}^{-1}$ for the agonist and $10^{4}~\mathrm{s}^{-1}$ for the self population, close to the maximum shown in that figure.

At $\nkp=10,\ \om=5$ the true median first activation time for the self population is $5.03\times10^{3}$ s
at $k_{\rm p}=1$ s$^{-1}$, or $\approx 67$ min at the $k_{\rm p}=1.25$ s$^{-1}$ implied
by $\tau_{\rm KPR}=8$ s. Eq.~(1) yields 1.74 s, or 1.39 s at $k_{\rm p}=1.25$ s$^{-1}$. Eq.~(5) gives
$5.03\times10^{3}$ s, the same 67 min to three digits, so the closed form recovers the
regime distinction that Eq.~(1) erases.

That hour is set by the rebinding rate as much as by $\koff$. Once retries dominate, the trigger rate is $\kappa=\sigma/\tau_{\rm c}$ [Eq.~(8)], and $\tau_{\rm c}$ contains $1/k_1$, so at $\nkp=10$, $\om=5$ and $\kp=1.25~\mathrm{s}^{-1}$ the exact median is about 11~min for $k_1=10^4\kp$, 67~min for $k_1=\kp$, 9.5~h for $k_1=\kp/10$ and 93~h for $k_1=\kp/100$. This does not contradict Table~\ref{tab:knobs}: rebinding is second order for the error of Eq.~(1), which lives in the first-attempt regime, but first order for the activation time once retries dominate. The setting of Ref.~\cite{morgan2023}, in which self ligands rebind at an effective $10^{4}~\mathrm{s}^{-1}$, lies at the fast end of this range.

Mempel, Henrickson and von Andrian showed by two-photon imaging that T cell
priming proceeds in three phases: over the first $8$~h T cells make multiple brief DC encounters, and over
the following $12$~h they form long-lasting stable
conjugates~\cite{mempel2004}.

Eq.~(1) would have self antigen trigger in under two seconds, making almost every contact a false positive. The exact answer of roughly an hour is
survivable through the brief contacts of phase~I. It becomes a real risk only in
the prolonged conjugates of phase~II, which is the regime in which the optimal
contact duration $\tau^{*}$ of Ref.~\cite{morgan2023} is finite rather than
infinite. We note that individual phase-II conjugates are themselves of order an
hour, so the margin here is one of the same order as the quantity being
protected, not a comfortable one.

The same factor carries over to the energetic budget. Ref.~\cite{morgan2023} counts the mean number $n_e$ of futile reactions, forward steps undone by dissociation, over a contact of duration $\tau$. Past relaxation $n_e$ grows linearly with $\tau$, so a contact duration set from the activation time of Eq.~(1) rather than the exact one understates the proofreading cost by the same factor of about 3000, up to an offset that is negligible once $\tau$ is large.

Shape collapse as seen in Table~\ref{tab:shape} is significant since a shape of $1$ is memory-less. So for ligands that a T cell should already reject (those that are fast to unbind with a high $\om$), proofreading still delivers specificity since the trigger rate is
suppressed by $r^{\nkp}$, $r=\kp/(\kp+\koff)$. However, this forfeits timing sharpness since discrimination survives but the safe contact duration threshold does not.

How these regimes bear on measured discrimination is not settled by the present model. The discriminatory power $\alpha\approx2$ measured for native TCRs~\cite{pettmann2021} is modest next to what a long proofreading chain could deliver, but Pettmann et al.\ fitted their own data with a chain of about 2.7 steps, whereas the larger depths $\nkp=4.5$--$11.3$ come from a light-gated chimeric receptor~\cite{britain2022}. Theory is also divided on how much reliability extra steps buy once stochasticity is included~\cite{kirby2023}; the answer depends on the time available to decide~\cite{xiao2024} and on the readout~\cite{li2024}, and adding a single negative feedback to a proofreading chain reproduces sensitivity, specificity and antagonism together~\cite{francois2013}. What the present results add is that longer chains also move the receptor into the regime where the first-passage time is set by rebinding statistics rather than by the ladder itself.

\section{Validation of simulations} 

Gillespie simulation is impractical at $\nkp=10,\ \om=5$ with
$n=10^4$ pairs each jumping at $\kp+\koff=6~\mathrm{s}^{-1}$ over a median of
$5\times10^3$~s, one replicate is $\sim\!3\times10^{8}$ jump events. We therefore simulate at the level of whole binding \emph{attempts}. The number of failed attempts before success is geometric with success probability $\sigma$ (Sec.~\ref{sec:closed}), and the step at which each failed attempt ends is multinomial over the $\nkp$ bound states, which fixes the total number of bound holding periods. The holding times then collapse into three Gamma draws (the time spent unbound, the time spent in bound states during failed attempts, and the successful traverse), so that one pair costs $O(\nkp)$ draws however many million attempts it makes.

Since no model is fitted, $R^2$ has a known ceiling. With midpoint plotting
positions $p_i=(i-\tfrac12)/N$ the total sum of squares is exactly $(N^2-1)/(12N)$. The expected
residual sum has a variance part $N/(6(N{+}1))$ and a bias part $(N-1)/\bigl(12N(N{+}1)\bigr)$ arising from the offset between $p_i$ and
$\mathbb{E}[U_{(i)}]=i/(N{+}1)$, so a model that is exactly right scores
\[
\mathbb{E}\left[R^{2}\right]=1-\frac{2N-1}{N^{2}-1}=0.999667
\qquad (N=6000).
\]
Table~\ref{tab:r2} gives the 25 values.

\begin{table}[!t]
\centering\small
\caption{$R^2$ between the exact law and the simulation. Largest vertical distance between the two
curves anywhere on the grid: $0.019$. The sampling standard deviation of $R^2$ at $N = 6000$ is $3.0  \times 10^{-4}$, comparable to the gap from unity. The residual sum is right-skewed, so $R^2$ itself has a long lower tail. The ceiling is
a mean rather than a bound, so individual cells scatter on both sides of it, with the
central 95\% interval for an exactly correct model lying in [0.9988, 0.9999].}
\label{tab:r2}
\begin{tabular}{@{}lrrrrr@{}}
\toprule
& $\om=1$ & $\om=2$ & $\om=3$ & $\om=4$ & $\om=5$ \\
\midrule
$\nkp=2$  & 0.99985 & 0.99887 & 0.99916 & 0.99992 & 0.99874 \\
$\nkp=4$  & 0.99983 & 0.99967 & 0.99987 & 0.99989 & 0.99989 \\
$\nkp=6$  & 0.99982 & 0.99985 & 0.99995 & 0.99957 & 0.99929 \\
$\nkp=8$  & 0.99986 & 0.99934 & 0.99993 & 0.99979 & 0.99899 \\
$\nkp=10$ & 0.99960 & 0.99987 & 0.99936 & 0.99970 & 0.99989 \\
\bottomrule
\end{tabular}
\end{table}

\section{Conclusion}
To leading order the relative error of Eq.~(1) is
\[
\min(\Lambda,1),\qquad \Lambda\;=\;\frac{(\kp+\koff)\,t_{q}}{\nkp+1},
\]
the Weibull-predicted activation time measured in units of $\nkp+1$ bound-state lifetimes. Eq.~(1) is therefore accurate only when $\Lambda\ll1$, and accurate to a tolerance $\epsilon$ roughly when $\Lambda\lesssim\epsilon$. $\nkp$ and $n$ set $t_{q}$ while $\om$ sets the lifetime, so neither alone determines whether Eq.~(1) can be used. Where it cannot, Eq.~(5) restores $\omega$ for the cost of a one-dimensional root find on the bracket $[t_q^{\rm W},t_q^{\rm U}]$. It reduces to Eq.~(1) when $\Lambda\ll1$, reproduces the shape collapse that Eq.~(1) cannot express, and tracks the exact median to within 2\% in 19 of the 25 cells of the grid and within 6.1\% in the worst.

\section*{Reproducibility}

Exact laws use uniformization (positive Poisson sums, full relative
accuracy down to $F\sim10^{-67}$ that are verified against exact rational arithmetic).
The density is $f(t)=\kp\left(\mathrm{e}^{Qt}\right)_{C_0,C_{\nkp-1}}$ for a
single pair, taken
from the same sweep rather than by differencing $F$. All figures and numbers come from the \texttt{fram} package (version~1.0.0, \url{https://pypi.org/project/fram/}, DOI~10.5281/zenodo.22999081), which accompanies this paper; it takes its name from the First Receptor Activation Model of Ref.~\cite{morgan2023}, which it reimplements and extends. Figures~\ref{fig:eq1grid} and~\ref{fig:pdfgrid} come from a set of 150\,000 simulated
contacts (25 cells $\times$ 6000). Figure~\ref{fig:sep} is computed from the exact law
alone, over a separate 63-point sweep. Figure~\ref{fig:newgrid} is drawn on the same 150\,000 simulated contacts (same seeds)
and the same exact curve as Figs.~\ref{fig:eq1grid} and~\ref{fig:pdfgrid}, so its
comparison with Fig.~\ref{fig:eq1grid} and Table~\ref{tab:r2} is paired: any difference
between the two figures comes from the formula rather than from resampling. Eq.~(5) rather
than Eq.~(4) is the form evaluated in code since the two agree to the order kept, but
Eq.~(4) differences two nearly equal incomplete gamma terms and loses most of its
significant figures when $\sigma$ is small. A suite of 465 test cases (162 test functions) validates the results against both the
exact CDF and the jump-by-jump Gillespie SSA~\cite{gillespie} where the latter is
affordable, using the same simulation machinery as in Refs.~\cite{wong2026,wong2_2026}.

\section*{Data availability statement}
The data that support the findings of this study are openly available at the following URL/DOI: \url{https://doi.org/10.5281/zenodo.22999081}.

\section*{Competing interests}
The authors declare no competing interests.

\section*{Author contributions}
T.M.: Conceptualization, methodology, and software. C.H.: Validation and formal analysis. K.W.: Investigation.


\section*{References}
\begin{thebibliography}{99}
\bibitem{irvine2002}
D.~J. Irvine, M.~A. Purbhoo, M.~Krogsgaard, and M.~M. Davis.
\newblock Direct observation of ligand recognition by T cells.
\newblock \emph{Nature}, 419(6909):845--849, 2002.

\bibitem{sykulev1996}
Y.~Sykulev, M.~Joo, I.~Vturina, T.~J. Tsomides, and H.~N. Eisen.
\newblock Evidence that a single peptide--MHC complex on a target cell can
  elicit a cytolytic T cell response.
\newblock \emph{Immunity}, 4(6):565--571, 1996.

\bibitem{purbhoo2004}
M.~A. Purbhoo, D.~J. Irvine, J.~B. Huppa, and M.~M. Davis.
\newblock T cell killing does not require the formation of a stable mature
  immunological synapse.
\newblock \emph{Nat. Immunol.}, 5(5):524--530, 2004.

\bibitem{smithgarvin2009}
J.~E. Smith-Garvin, G.~A. Koretzky, and M.~S. Jordan.
\newblock T cell activation.
\newblock \emph{Annu. Rev. Immunol.}, 27:591--619, 2009.

\bibitem{courtney2018}
A.~H. Courtney, W.-L. Lo, and A.~Weiss.
\newblock TCR signaling: mechanisms of initiation and propagation.
\newblock \emph{Trends Biochem. Sci.}, 43(2):108--123, 2018.

\bibitem{gaud2018}
G.~Gaud, R.~Lesourne, and P.~E. Love.
\newblock Regulatory mechanisms in T cell receptor signalling.
\newblock \emph{Nat. Rev. Immunol.}, 18(8):485--497, 2018.

\bibitem{matsui1994}
K.~Matsui, J.~J. Boniface, P.~Steffner, P.~A. Reay, and M.~M. Davis.
\newblock Kinetics of T-cell receptor binding to peptide/I-E$^k$ complexes:
  correlation of the dissociation rate with T-cell responsiveness.
\newblock \emph{Proc. Natl. Acad. Sci. USA}, 91(26):12862--12866, 1994.

\bibitem{aleksic2010}
M.~Aleksic, O.~Dushek, H.~Zhang, E.~Shenderov, J.-L. Chen, V.~Cerundolo,
  D.~Coombs, and P.~A. van~der Merwe.
\newblock Dependence of T cell antigen recognition on T cell receptor--peptide
  MHC confinement time.
\newblock \emph{Immunity}, 32(2):163--174, 2010.

\bibitem{pettmann2021}
J.~Pettmann, A.~Huhn, E.~Abu~Shah, M.~A. Kutuzov, D.~B. Wilson, M.~L. Dustin,
  S.~J. Davis, P.~A. van~der Merwe, and O.~Dushek.
\newblock The discriminatory power of the T cell receptor.
\newblock \emph{eLife}, 10:e67092, 2021.

\bibitem{hopfield1974}
J.~J. Hopfield.
\newblock Kinetic proofreading: a new mechanism for reducing errors in
  biosynthetic processes requiring high specificity.
\newblock \emph{Proc. Natl. Acad. Sci. USA}, 71(10):4135--4139, 1974.

\bibitem{ninio1975}
J.~Ninio.
\newblock Kinetic amplification of enzyme discrimination.
\newblock \emph{Biochimie}, 57(5):587--595, 1975.

\bibitem{mckeithan1995}
T.~W. McKeithan.
\newblock Kinetic proofreading in T-cell receptor signal transduction.
\newblock \emph{Proc. Natl. Acad. Sci. USA}, 92(11):5042--5046, 1995.

\bibitem{yousefi2019}
O.~S. Yousefi, M.~G\"unther, M.~H\"orner, J.~Chalupsky, M.~Wess,
  S.~M. Brandl, R.~W. Smith, C.~Fleck, T.~Kunkel, M.~D. Zurbriggen, T.~H\"ofer,
  W.~Weber, and W.~W.~A. Schamel.
\newblock Optogenetic control shows that kinetic proofreading regulates the
  activity of the T cell receptor.
\newblock \emph{eLife}, 8:e42475, 2019.

\bibitem{tischer2019}
D.~K. Tischer and O.~D. Weiner.
\newblock Light-based tuning of ligand half-life supports kinetic proofreading
  model of T cell signaling.
\newblock \emph{eLife}, 8:e42498, 2019.

\bibitem{britain2022}
D.~M. Britain, J.~P. Town, and O.~D. Weiner.
\newblock Progressive enhancement of kinetic proofreading in T cell antigen
  discrimination from receptor activation to DAG generation.
\newblock \emph{eLife}, 11:e75263, 2022.

\bibitem{li2024}
X.~Li and T.~Chou.
\newblock Reliable ligand discrimination in stochastic multistep kinetic
  proofreading: first passage time vs. product counting strategies.
\newblock \emph{PLoS Comput. Biol.}, 20(6):e1012183, 2024.

\bibitem{morgan2023}
J.~Morgan and A.~E. Lindsay.
\newblock Modulation of antigen discrimination by duration of immune contacts in
  a kinetic proofreading model of T cell activation with extreme statistics.
\newblock \emph{PLoS Comput. Biol.}, 19(8):e1011216, 2023.
\newblock \href{https://doi.org/10.1371/journal.pcbi.1011216}{doi:10.1371/journal.pcbi.1011216}.

\bibitem{bel2010}
G.~Bel, B.~Munsky, and I.~Nemenman.
\newblock The simplicity of completion time distributions for common complex
  biochemical processes.
\newblock \emph{Phys. Biol.}, 7(1):016003, 2010.

\bibitem{munsky2009}
B.~Munsky, I.~Nemenman, and G.~Bel.
\newblock Specificity and completion time distributions of biochemical
  processes.
\newblock \emph{J. Chem. Phys.}, 131(23):235103, 2009.

\bibitem{kirby2023}
D.~Kirby and A.~Zilman.
\newblock Proofreading does not result in more reliable ligand discrimination
  in receptor signaling due to its inherent stochasticity.
\newblock \emph{Proc. Natl. Acad. Sci. USA}, 120(21):e2212795120, 2023.

\bibitem{xiao2024}
F.~Xiao and V.~Galstyan.
\newblock With the leisure of time, kinetic proofreading can still perform
  reliable ligand discrimination.
\newblock \emph{Proc. Natl. Acad. Sci. USA}, 121(32):e2320118121, 2024.

\bibitem{lawley2020networks}
S.~D. Lawley.
\newblock Extreme first-passage times for random walks on networks.
\newblock \emph{Phys. Rev. E}, 102(6):062118, 2020.

\bibitem{madrid2020}
J.~B. Madrid and S.~D. Lawley.
\newblock Competition between slow and fast regimes for extreme first passage
  times of diffusion.
\newblock \emph{J. Phys. A: Math. Theor.}, 53(33):335002, 2020.

\bibitem{maclaurin2025}
J.~MacLaurin and J.~Newby.
\newblock Extreme first passage times for populations of identical rare
  events.
\newblock \emph{SIAM J. Appl. Math.}, 85(1):109--142, 2025.

\bibitem{grebenkov2020}
D.~S. Grebenkov, R.~Metzler, and G.~Oshanin.
\newblock From single-particle stochastic kinetics to macroscopic reaction
  rates: fastest first-passage time of {$N$} random walkers.
\newblock \emph{New J. Phys.}, 22(10):103004, 2020.

\bibitem{lawley2024}
S.~D. Lawley.
\newblock Competition of many searchers.
\newblock In D.~S. Grebenkov, R.~Metzler, and G.~Oshanin, editors,
  \emph{Target Search Problems}, pages 281--303. Springer, Cham, 2024.

\bibitem{kalashnikov1997}
V.~Kalashnikov.
\newblock \emph{Geometric Sums: Bounds for Rare Events with Applications}.
\newblock Kluwer Academic, Dordrecht, 1997.

\bibitem{mempel2004}
T.~R. Mempel, S.~E. Henrickson, and U.~H. von Andrian.
\newblock T-cell priming by dendritic cells in lymph nodes occurs in three
  distinct phases.
\newblock \emph{Nature}, 427(6970):154--159, 2004.

\bibitem{francois2013}
P.~François, G.~Voisinne, E.~D. Siggia, G.~Altan-Bonnet, and M.~Vergassola.
\newblock Phenotypic model for early T-cell activation displaying sensitivity,
  specificity, and antagonism.
\newblock \emph{Proc. Natl. Acad. Sci. USA}, 110(10):E888--E897, 2013.

\bibitem{gillespie}
D. T. Gillespie.
\newblock A general method for numerically simulating the stochastic time
evolution of coupled chemical reactions.
\newblock \emph{J. Comput. Phys.}, 22(4):403--434, 1976.

\bibitem{wong2026}
K.~Wong, T.~Mourier, and C.~Hopkinson.
\newblock Post-transcriptional regulation of stochastic gene expression conditioned on large deviations.
\newblock \emph{arXiv}:2607.17004, 2026.

\bibitem{wong2_2026}
K.~Wong, T.~Mourier, S.~Inaba, C.~Hopkinson, and R.~Kulkarni.
\newblock Approximate analytical protein distributions for the three-stage model of stochastic gene expression.
\newblock \emph{arXiv}:2608.08916, 2026.
\end{thebibliography}
\end{document}